\documentclass[11pt]{article}
\usepackage{moriond,epsfig}

\def\Journal#1#2#3#4{{#1} {\bf #2}, #3 (#4)}

\def\be{\begin{equation}}
\def\ee{\end{equation}}
\def\bea{\begin{eqnarray}}
\def\eea{\end{eqnarray}}

\begin{document}
\vspace*{4cm}
\title{JET MEASUREMENTS AND EXTRACTION OF THE STRONG COUPLING CONSTANT AT HERA}

\author{ A.A. SAVIN \footnote{On behalf of the H1 and ZEUS collaborations}}

\address{
University of Wisconsin-Madison \\
1150 University Ave., Madison WI 5370-1390, USA}

\maketitle\abstracts{
Results on jet measurements in neutral current deep inelastic scattering at HERA
are presented. The low-$x_{Bj}$ and low-$Q^2$ region is explicitly investigated using
forward jet production and the azimuthal asymmetry between jets in dijet 
production. Recent results on the determination of the strong coupling
constant, $\alpha_s(M_Z)$, are discussed.}

\section{Introduction}

Precise measurements of jet production in photoproduction
(PHP) and deep inelastic scattering (DIS) demonstrate good understanding of
underlying physics and great success of the QCD-based theoretical calculations.
In PHP, resolved photon process, where the photon fluctuates into a source of
partons, one of which takes part in the hard scatter, plays an important role. 
Such a process in  $\gamma p$ scattering should be quite similar to  
$hh$ collisions. With
increasing virtuality of the photon the resolved contribution decreases
rapidly. Recent studies in DIS inclusive jet and dijet production
\cite{H12jets,ZEUS2jets} 
showed however, that even at moderate $Q^2$, 10-50~GeV$^2$, the contribution
from resolved photons is still important. Moreover this
contribution is essential in the region of low $x_{Bj}$ and low $Q^2$,
where the DGLAP-based calculations are expected to fail and the BFKL dynamics 
should be more prononced.

\section{Jet production in the low-$x_{Bj}$, low-$Q^2$ region}

Jet production in the forward, proton direction, pseudorapidity region 
of $\gamma^* p$ scattering was a long time ago suggested as a good
testing ground to study effects of the BFKL dynamics~\cite{Mueller}. 
In this region, the production
of high-$E_T$ jets via DGLAP is suppressed because
of the $k_T$ ordering in the $Q^2$ evolution. Such an ordering requires that 
the highest $E_T$ jet (corresponding to the highest $k_T$-parton) 
should come from the hard interaction.
In dijet production, the two highest-$E_T$ jets should be always back-to-back,
compensated both in the energy and in angles (in the leading order of
$\alpha_s$). It is not necessarily true in 
the case of BFKL, where high $k_T$ partons can also appear quite close to the 
proton and will carry
a big portion of its longitudinal momentum, thus creating high-$E_T$
jets in the forward direction. A strong azimuthal correlation in dijet 
production is also not required in the BFKL case.

 \begin{figure}[htbp]
\begin{center}
\includegraphics[width=65mm]{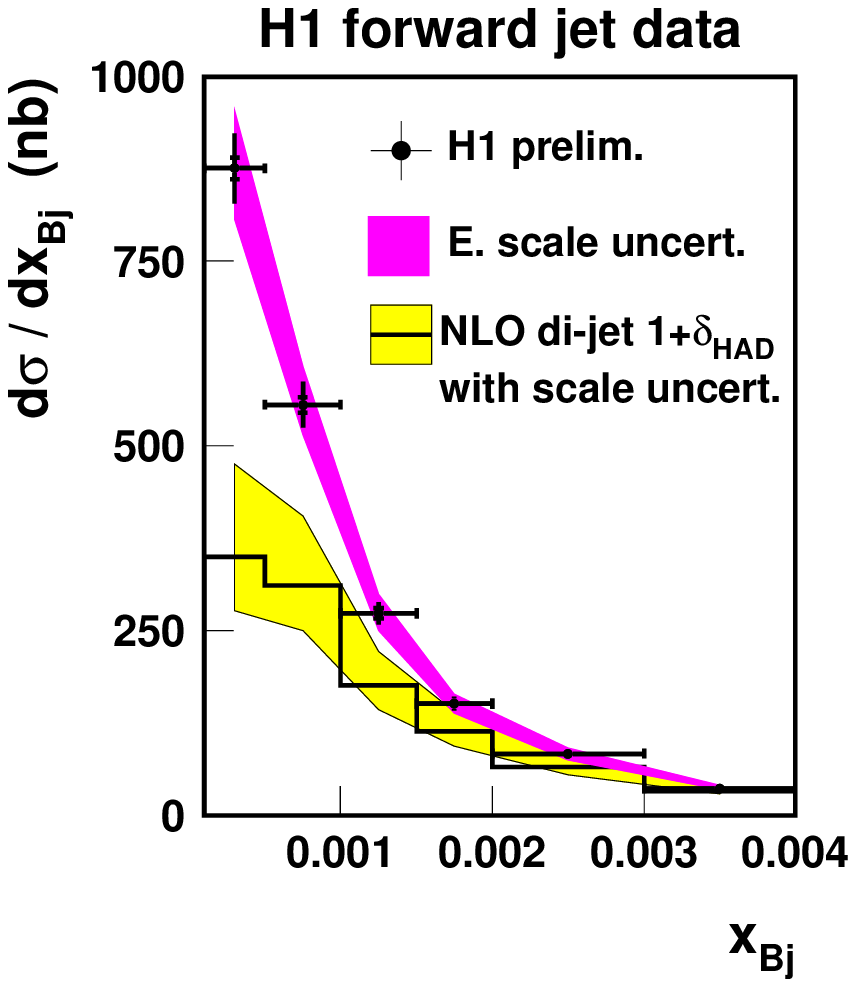}
\includegraphics[width=65mm]{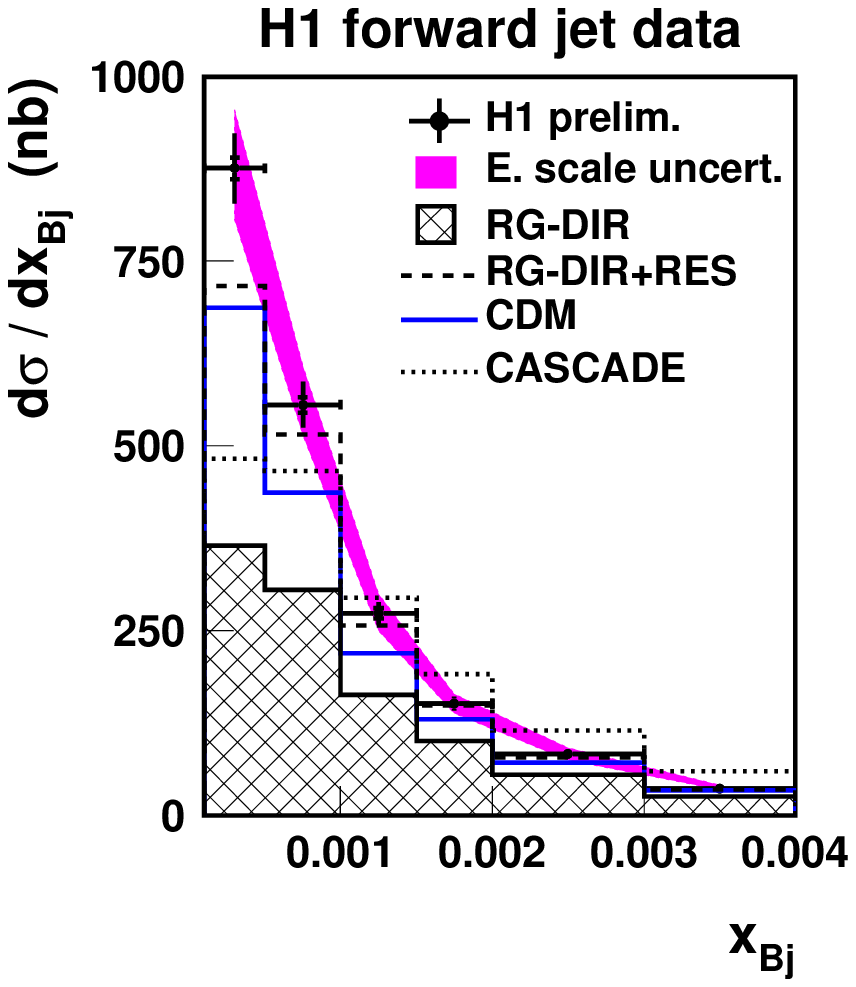}
\caption{The hadron level cross section for inclusive forward jet production as
a function of $x_{Bj}$ compared to predictions of NLO calculations (lhs plot)
and different Monte Carlo models (rhs plot) \label{fig:H1fj}}
\end{center}
\end{figure}

\begin{figure}[htbp]
\begin{center}
\includegraphics[width=80mm]{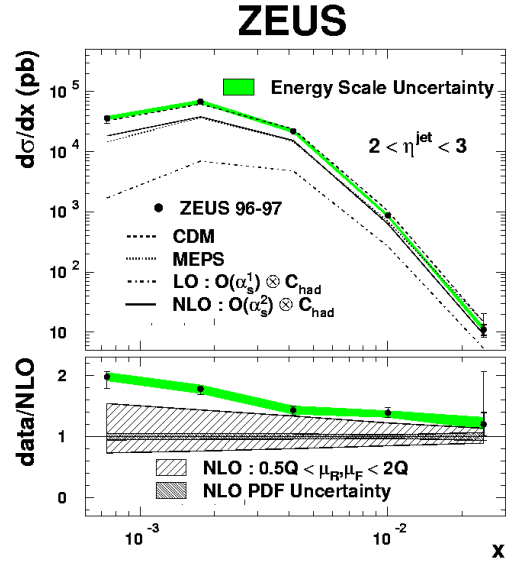}
\caption{Differential cross section for the forward jets production as a
function of $x_{Bj}$ measured by ZEUS~\protect\cite{ZEUSfj}\label{fig:ZEUSfj}.
Predictions of different QCD models and calculations explained in
the text are shown.}
\end{center}
\end{figure}

\begin{figure}[htbp]
\begin{center}
\includegraphics[width=95mm]{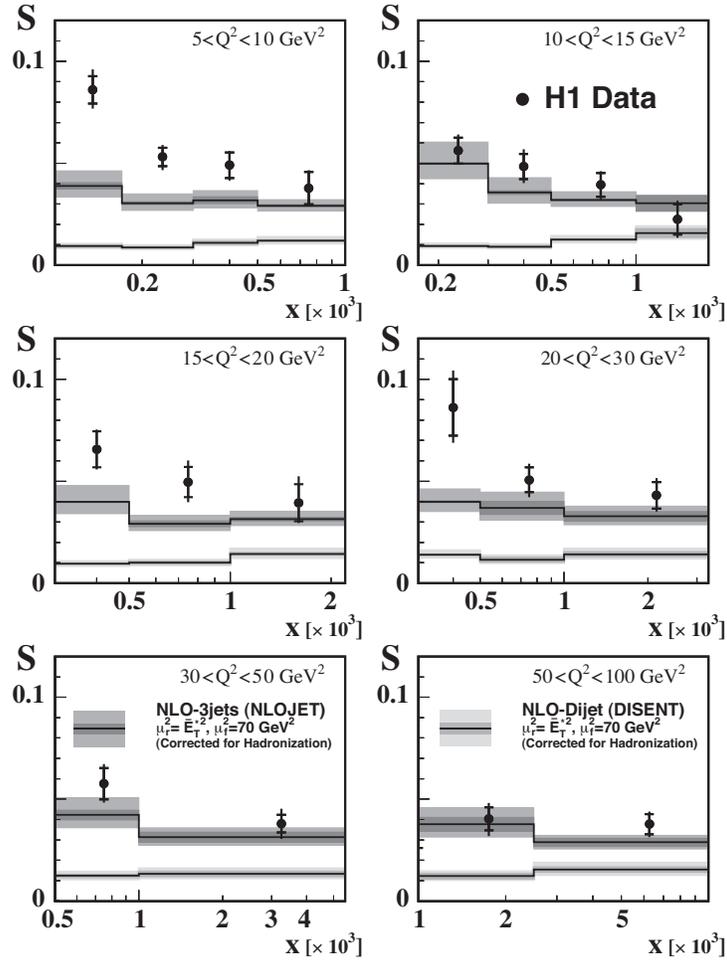}
\caption{Ratio $S$ of the number of dijet events with a small
azimuthal jet separation ($\Delta \phi < 120 ^\circ$) between the two
highest $E_T$ jets with respect to the total number of inclusive
dijets events, as a function of $x_{Bj}$ and $Q^2$ \label{fig:H1aa}}
\end{center}
\end{figure}

Figure ~\ref{fig:H1fj} shows the H1 results~\cite{H1fj} on inclusive
forward jet production at $5<Q^2<85$ GeV$^2$ and $0.0001<x<0.004$.
The jets are defined using the $k_T$ algorithm in its inclusive mode
and by requiring in the laboratory frame $p_{T,jet}>3.5$~GeV, $7^\circ <
\Theta_{jet} < 20^\circ$. To suppress the phase space for DGLAP evolution
the requirements $0.5<p^2_{T,jet}/Q^2<5$ and $E_{jet}/E_{proton}>0.035$ were
applied.
The cross section as a function of $x_{Bj}$ compared to the predictions of the 
NLO QCD calculation from DISENT shows that the calculation does 
not describe the data at low $x_{Bj}$.
The rhs plot shows comparison with predictions of the RAPGAP
model, where including a resolved photon contribution a reasonable
description of the data is observed. The CCFM-model (CASCADE) 
predicts a somewhat different
shape for the
$x_{Bj}$ distribution, which results in a comparatively poor description of the
data.

It is interesting to compare the resolved photon
effects with results of NNLO contribution, however
current technics does not allow
to perform such a calculation. 
To check the NNLO effects ZEUS
restricted the inclusive forward jet measurement\cite{ZEUSfj} by requiring 
at least one jet to be in the forward region together with significant
hadronic activities in the backward one. Such a requirement suppresses the
$\alpha_s^0$ contributions. The NLO calculation in this case can be generated at next order of $\alpha_s$ 
compare to the H1 inclusive measurement.

The ZEUS cross section measurement is presented in Fig.~\ref{fig:ZEUSfj} for
$Q^2>25$ GeV$^2$ and $E_T^{jet}>6$~GeV. The measurement is performed in the very
forward region
$2<\eta^{jet}<3$. The DGLAP contribution is suppressed by requiring
$0.5<(E_T^{jet})^2/Q^2<2$. 
Single jet production is also suppressed using
the total angle of the final hadronic system $\gamma_h$. The $cos(\gamma_h)$
had to be less then 0, thus requiring a sufficient backwards activities together
with a jet in the forward direction.

The NLO still does not describe the data, although 
taking in account the size of the theoretical uncertainties 
the disagreement is not so dramatic. 
The difference between DGLAP and BFKL dynamics can be
also studied using different parton-shower models used in the Monte Carlo
programs. The  
DGLAP-like MEPS (LEPTO) fails to describe the data at low $x_{Bj}$. The  
BFKL-like CDM (ARIADNE) describes the data quite well.

Using an inclusive dijet sample, H1 has measured azimuthal correlations between jets
\cite{H1aa}
in quite restricted phase space: $5<Q^2<100$~GeV$^2$, $10^{-4}<x<10^{-2}$ . The variable $S$ represents the ratio of 
number of events with azimuthal
difference between two highest $E_T$ jets, $\Delta \phi$, less then 120$^\circ$ to the total number
of dijet events; see Fig.~\ref{fig:H1aa}. The NLO calculation predicts that 
the $S$ should be very close to 0 in all the bins of the measurement. The
NLO QCD calculation for trijet production prediction in more  compatible 
with the data, but still 
insufficient to describe data at low $x_{Bj}$ and low $Q^2$. 

Comparison of the recent HERA jet data with
predictions of QCD-based models demonstrates, that to describe the data
in low-$x_{Bj}$, low-$Q^2$ region one needs or to include the contribution
from the resolved
photon process, or to perform calculation extended to next orders of
$\alpha_s$, or to develop a new BFKL-based, 
QCD calculation.

\section{Inclusive jet production and determination of $\alpha_s(M_Z)$}

\begin{figure}[p]
\begin{center}
\includegraphics[width=75mm]{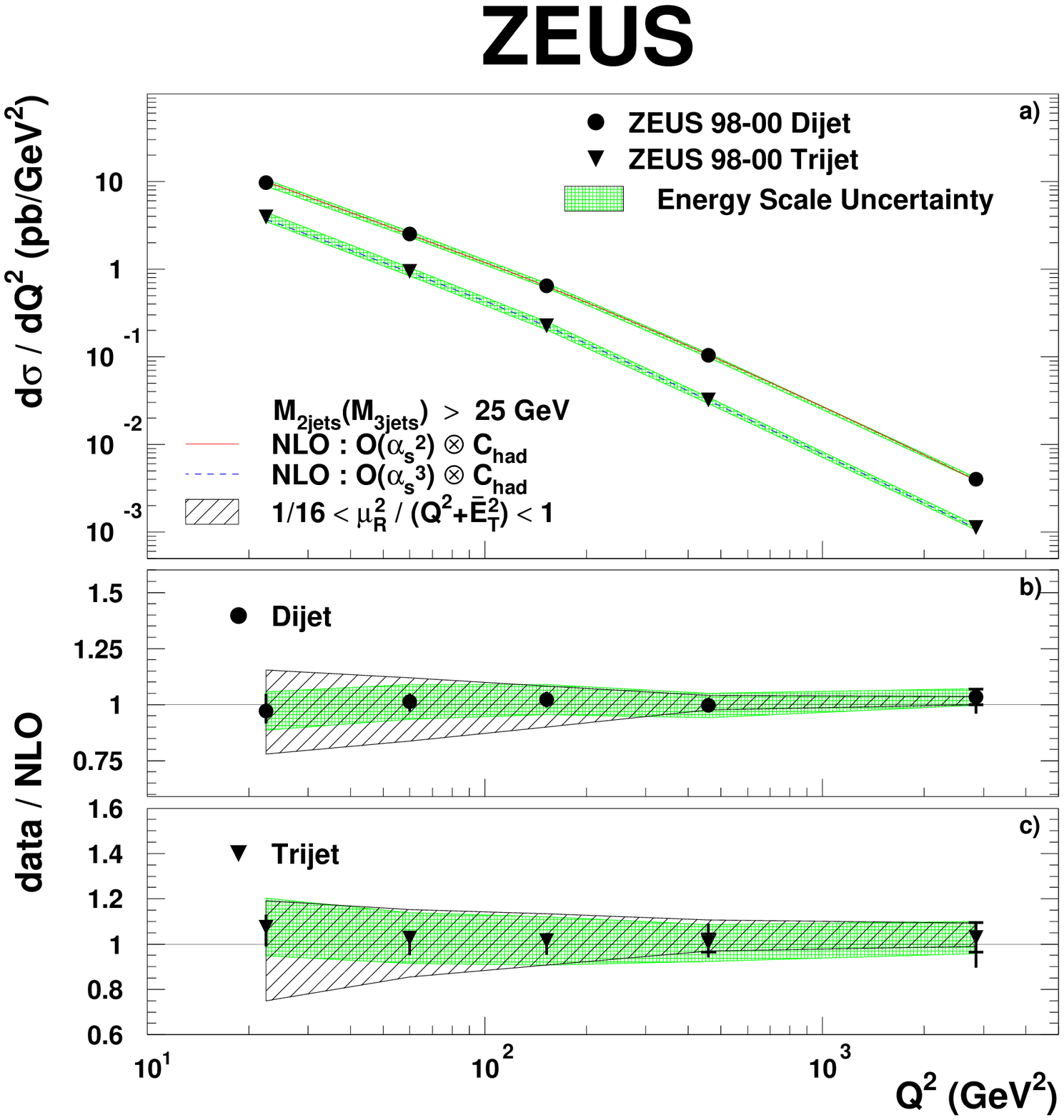}
\includegraphics[width=75mm]{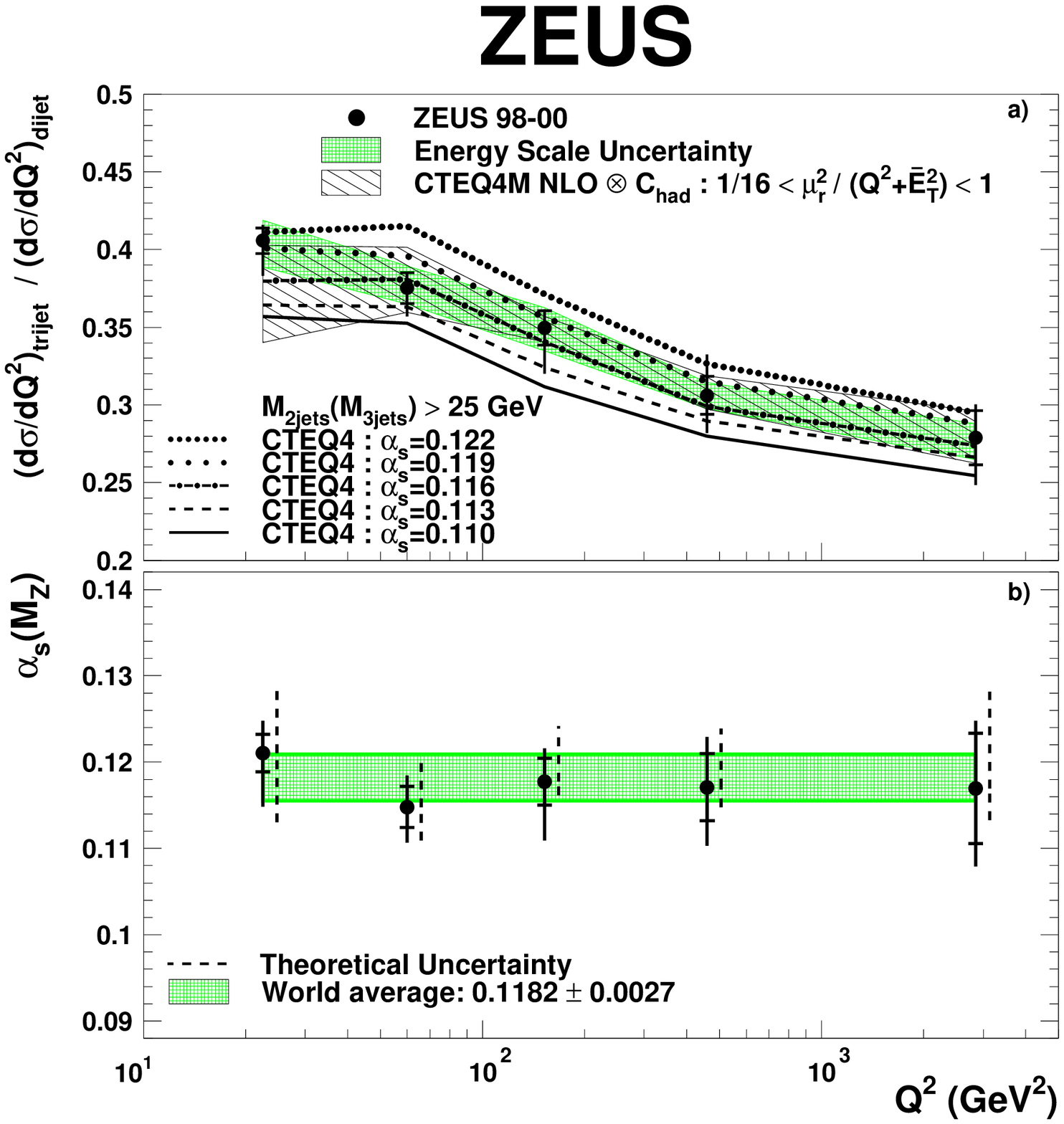}
\caption{The inclusive dijet and trijet cross sections as a function of $Q^2$.
The predictions of perturbative QCD in next-to-leading order are compared to the
data. Rhs plot shows the ratio of inclusive trijet to dijet cross sections and
determined $\alpha_s(M_Z)$ values in different regions of $Q^2$ 
\label{fig:ZEUS23j}}
\end{center}
\end{figure}

Apart from the above mentioned region of phase space, where
DGLAP-based calculations are not expected to work 
and indeed do not describe the data, the
agreement between the data and predictions is good. Results of recent
inclusive dijet and trijet measurements by ZEUS\cite{ZEUSmultijets}
are presented in 
Fig.~\ref{fig:ZEUS23j}. The NLOJET calculation describes the
data well, as shown in the lhs plot, thus making possible an extraction
of the $\alpha_s(M_Z)$. The ratio of the dijet to trijet cross sections 
as a function of $Q^2$ is shown in the rhs plot together with QCD 
predictions based on different values of $\alpha_s(M_Z)$. It can be seen that the
measurement is very sensitive to the $\alpha_s(M_Z)$ value. Also both the
experimental and theoretical uncertainties are reduced in the ratio 
compared to the cross
section measurement, since some of the uncertainties cancel.

\begin{figure}[htbp]
\begin{center}
\includegraphics[width=80mm]{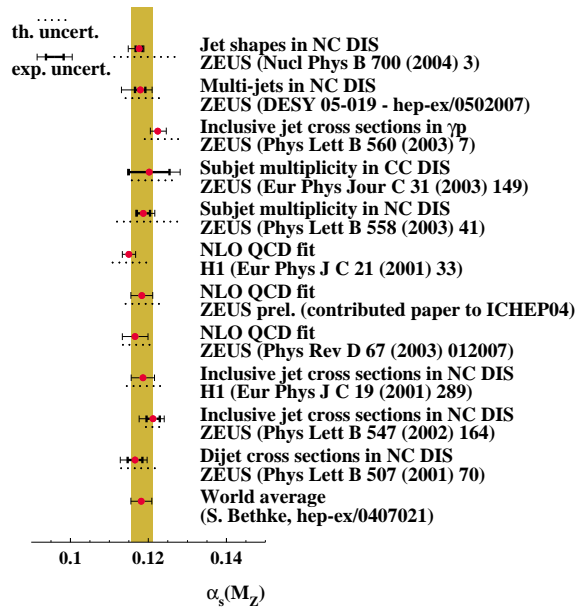}
\caption{Summary of $\alpha_s(M_Z)$ measurements at HERA \label{fig:alpha}}
\end{center}
\end{figure}

The extracted values of $\alpha_s (M_Z)$ in each $Q^2$ bin are also shown in the
bottom part of the rhs plot of the Fig.~\ref{fig:ZEUS23j}. 
A $\chi^2$-fit to these
values gives an average $\alpha_s(M_Z)$ value, that is shown in the 
Fig.~\ref{fig:alpha} together with the previous $\alpha_s(M_Z)$ determinations from HERA and 
compared to the world average $\alpha_s(M_Z)$ value.

\end{document}